\documentclass[aps,prd,showpacs]{revtex4}
\usepackage{graphicx}
\usepackage{psfig}
\usepackage{revsymb}
\begin{document}
\title{Classical Trajectories in Rindler Space and Restricted
Structure of Phase Space with Non-Hermitian $PT$-Symmetric Hamiltonian}
\author{\footnotesize Soma Mitra$^{a,1}$ and Somenath Chakrabarty$^{a,2}$}
\affiliation{$^a$Department of Physics, Visva-Bharati, Santiniketan 731235, 
India\\
$^1$somaphysics@gmail.com\\
$^2$somenath.chakrabarty@visva-bharati.ac.in
}
\begin{abstract}
The nature of single particle classical phase space trajectories in
Rindler space with non-hermitian $PT$-symmetric Hamiltonian
have been studied both in the relativistic as well as in the non-relativistic scenarios. It has been shown that 
in the relativistic scenario, both  positional coordinates and the corresponding canonical momenta are real in
nature and diverges with time. Whereas the phase space trajectories are a set of hyperbolas in Rindler space. On the
other hand in the non-relativistic approximation the spatial coordinates are complex in nature, whereas the
corresponding canonical momenta of the particle are purely imaginary. In this case the phase space trajectories are
quite simple in nature. But the spatial coordinates are restricted in the negative region only.
\end{abstract}
\maketitle
\section{Introduction}
Exactly like the Lorentz transformations of space time coordinates in
the inertial frame \cite{R1,R2}, the Rindler coordinate
transformations are for the uniformly accelerated frame of
references \cite{R3,R31,R32,R4,R5,R6,R7}. 
From the references \cite{R3,R31,R32,R4,R5,R6,R7}, it can very easily be
shown that the Rindler 
coordinate transformations are given by: 
\begin{eqnarray}
ct&=&\left (\frac{c^2}{\alpha}+x^\prime\right )\sinh\left (\frac{\alpha t^\prime}
{c}\right ) ~~{\rm{and}}~~ \nonumber \\
x&=&\left (\frac{c^2}{\alpha}+x^\prime\right )\cosh\left (\frac{\alpha t^\prime}
{c}\right ) 
\end{eqnarray}
Hence it is a matter of simple algebra to prove that the  inverse
transformations are given by:
\begin{equation}
ct^\prime=\frac{c^2}{2\alpha}\ln\left (\frac{x+ct}{x-ct}\right )
~~{\rm{and}}~~ x^\prime=(x^2-(ct)^2)^{1/2}-\frac{c^2}{\alpha}
\end{equation}
Here $\alpha$ indicates the uniform acceleration of the frame. Hence it can
very easily be shown from eqns.(1) and (2) that the square of the
four-line element changes from
\begin{eqnarray}
ds^2&=&d(ct)^2-dx^2-dy^2-dz^2 ~~{\rm{to}}~~\nonumber \\ ds^2&=&\left
(1+\frac{\alpha x^\prime}{c^2}\right)^2d(ct^\prime)^2-{dx^\prime}^2
-{dy^\prime}^2-{dz^\prime}^2
\end{eqnarray}
where the former line element is in the Minkowski  space.
Hence the metric in the Rindler space can be written as
\begin{equation}
g^{\mu\nu}={\rm{diag}}\left (\left (1+\frac{\alpha x}{c^2}\right
)^2,-1,-1,-1\right )
\end{equation}
whereas in the Minkowski space-time we have the usual form
\begin{equation}
g^{\mu\nu}={\rm{diag}}(+1, -1, -1, -1)
\end{equation}
It is therefore quite obvious that the Rindler space is also flat. The only difference from the Minkowski space is
that the frame of the observer is moving with uniform acceleration.
It has been noticed from the literature survey, 
that the principle of equivalence plays an important role in obtaining the
Rindler coordinates in the uniformly accelerated frame of reference. 
According to this principle an accelerated frame in absence
of gravity is equivalent to a frame at rest in presence of a
gravity. Therefore in the present scenario, $\alpha$ may be treated 
to be the strength of 
constant gravitational field for a frame at rest.

Now from the relativistic dynamics of special theory
of relativity \cite{R1}, the action integral is given by
\begin{equation}
S=-\alpha_0 c \int_a^b ds\equiv \int_a^b Ldt
\end{equation}
where $\alpha_0=-m_0 c$ \cite{R1} and $m_0$ is  the
rest mass of the particle and $c$ is the speed of light in vacuum.
The Lagrangian of the particle may be written as
\begin{equation}
L=-m_0c^2\left [\left ( 1+\frac{\alpha x}{c^2}\right )^2 -\frac{v^2}{c^2}
\right ]^{1/2}
\end{equation}
where $\vec v$ is the three velocity vector. Hence the three 
momentum of the particle is given by
\begin{equation}
\vec p=\frac{\partial L}{\partial \vec v}, ~~ {\rm{or}}
\end{equation}
\begin{equation}
\vec p=\frac{m_0\vec v}{\left [ \left (1+\frac{\alpha x}{c^2} \right )^2
-\frac{v^2}{c^2} \right ]^{1/2}}
\end{equation}
Then from the definition, the Hamiltonian of the particle may be written as
\begin{equation}
H=\vec p.\vec v-L ~~ {\rm{or}}
\end{equation}
\begin{equation}
H=m_0c^2 \left (1+\frac{\alpha x}{c^2}\right ) \left (1+
\frac{p^2}{m_0^2c^2}\right )^{1/2}
\end{equation}
Hence it can very easily be shown that in the non-relativistic approximation, the Hamiltonian is given by 
$$
H=\left (1+\frac{\alpha x}{c^2}\right ) \left (\frac{p^2}{2m_0}+m_0c^2 \right ) \eqno(11a)
$$
In the classical level, the quantities $H$, $x$ and $p$ are treated
as dynamical variables. 
Further, it can very easily be verified that in the quantum mechanical scenario where these quantities are
considered to be
operators, the Hamiltonian $H$ is not
hermitian. However the energy eigen spectrum for the Schr\"odinger equation has been observed to be real 
\cite{R71}. This is
found to be solely because of the fact that $H$ is $PT$-invariant. Now it is well know 
that $PxP^{-1}=-x$, $P p P^{-1}=-p$, whereas, $TpT^{-1}=-p$ and 
$P\alpha P^{-1}=-\alpha$ 
but $T\alpha T^{-1}=\alpha$, therefore it is a matter of simple
algebra to show that $PT~H~(PT)^{-1}=H^{PT}=H$. As has been shown by several
authors \cite{R10} that if $H$ is $PT$-invariant, then the energy
eigen values  will 
be real. Here $P$ and $T$ are respectively the parity and the time
reversal operators. Further if the Hamiltonian is $PT$ symmetric,
then $H$ and $PT$ should have common eigen states. In \cite{R71} we
have noticed that the solution of the Schr$\ddot{\rm{o}}$dinger
equation is obtained in terms of the variable  $u=1+\alpha x/c^2$,
which is $PT$-symmetric. Hence any function, e.g., 
Whittaker function $M_{k,\mu}(u)$ or Associated Laguerre function 
$L_m^n(u)$, the solution of the Schr\"odinger equation  are $PT$-symmetric. These polynomials are also the eigen
functions of the operator $PT$. 
Of course with the replacement of hermiticity
of the Hamiltonian with the $PT$-symmetry, we have not
discarded the important quantum mechanical key features of the system 
described by this Hamiltonian and also kept the canonical
quantization rule invariant, i.e., $TiT^{-1}=-i$. 
This point was
also discussed in an elaborate manner in reference \cite{R10} and in some of  the references cited there. 

In this article we have investigated the time evolution for both the space and the momentum coordinates of the
particle moving in Rindler space. We have considered both the relativistic and the non-relativistic form of the
Rindler Hamiltonian (eqns.(11) and (11a) respectively). Hence we
shall also obtain the classical phase space trajectories
for the particle in the Rindler space. We have noticed that in the relativistic scenario, both the spatial
and the momentum coordinates are real in nature and diverge as $t
\longrightarrow \infty$. For both the variables the 
time dependencies are extremely simple. Hence we have obtained classical trajectories $p(x)$ by eliminating
the time dependent part. 

However, in the non-relativistic approximation, the spatial coordinates are quite complex in nature, whereas the momentum
coordinates are purely imaginary. Since the mathematical form of the
phase space trajectories are quite complicated,
we have
obtained $p(x)$ numerically in the non-relativistic scenario. 

In the first part of this article, we have considered the relativistic picture and obtained the phase space
trajectories, whereas in the second part, the 
classical phase space structure is obtained for non-relativistic case.
To the best of our knowledge such studies have not been done before

\section{Relativistic Picture}
The classical Hamilton's equation of motion for the particle is 
given by \cite{R61}
\begin{equation}
\dot{x}=[H,x]_{p.x} ~~{\rm{and}}~~~ \dot{p}=[H,p]_{p,x}
\end{equation}
where $[H,f]_{p,x}$ 
is the Poisson bracket and is defined  by \cite{R61}
\begin{equation}
[f,g]_{p,x}=\frac{\partial f}{\partial p}\frac{\partial g}{\partial
x} -\frac{\partial f}{\partial x}\frac{\partial g}{\partial p}
\end{equation}
In this case $f=x$ or $p$. 
In eqn.(12) the dots indicate the derivative with respect to time. Now
using the relativistic version of Rindler Hamiltonian from eqn.(11), the explicit form of the
equations of motion are given by
\begin{equation}
\dot{x}=\left (1+\frac{\alpha x}{c^2}\right ) \frac{pc^2} {(p^2c^2+
m_0^2c^4)^{1/2}} ~~{\rm{and}}~~ \dot{p}=-\frac{\alpha}{c} (p^2 c^2+
m_0^2c^4)^{1/2}
\end{equation}
The parametric form of expressions for $x$ and $p$ represent the time evolution of
spatial coordinate and the corresponding canonical momentum. The
analytical expressions for time evolution for both the quantities
can be obtained
after integrating these coupled equations and are given by
\begin{equation}
x=\frac{c^2}{\alpha} [C_0 \cosh(\omega t-\phi)-1] ~~{\rm{and}}~~ 
p=-m_0c
~\sinh(\omega t-\phi)
\end{equation}
where $C_0$ and $\phi$ are the integration constants, which are real in
nature and $\omega=\alpha /c$ 
is the
frequency defined for some kind of quanta in \cite{R71}. Hence
eliminating the time coordinate, we can write
\begin{equation}
\left ( 1+\frac{\alpha x}{c^2}\right )^2\frac{1}{C_0^2}
-\frac{p^2}{m_0^2c^2}=1
\end{equation}
This is the mathematical form of the set of
classical trajectories of the particle in
the phase space. Or in other wards,
these set of hyperbolas are the classical trajectories of the particle
in the Rindler space. This is consistent with the hyperbolic motion of the
particle in a uniformly accelerated frame.
These set of hyperbolic equations can also be written as 
\begin{equation}
p^2= m_0^2c^2 \left (\frac{2\alpha x}{c^2}\right )
\left ( 1+\frac{x\omega}{2c}\right )
\end{equation}
It is quite obvious from the parametric form of the variation of $x$
and $p$ with time that both the quantities are unbound. This is also
reflected from the nature of phase space trajectories as shown in
fig.(1) for the scaled $x$ and $p$.  The scaling factors are
$\alpha/c^2$ for $x$ and $(m_0c)^{-1}$ for $p$. For the sake of
illustration, we have chosen the arbitrary constant $C_0=1$. 
In this figure
we have also taken both the scaling factors identically 
equal to unity. Then obviously eqn.(16) reduces to 
\[
(x+1)^2-p^2=1
\]
We shall get the
other set of trajectories by choosing different values for the scaling
factors. It is obvious that in this case the centre of the hyperbola 
is at $(-1, 0)$.
Therefore with the increase of $\alpha$, the centre $\longrightarrow
(0,0)$. Further the vertices for this particular hyperbolic curve are at
$(0,0)$ and  $(-2,0)$. The second one is in scaled form. Therefore for the gravitational field 
$\alpha$ large enough, both the vertices coincide at the centre
$(0,0)$. It is also obvious that for very large values of
$\alpha$, these two curves touch each other at $(0,0)$.
We have therefore noticed that the phase space trajectories are unbound and consistent with the motion of the
particle in Rindler space.

\section{Non-relativistic Picture}
We next consider the non-relativistic form of Rindler Hamiltonian given by eqn.(11a). Now following 
eqn.(12), the equations of motion for the particle in Rindler space
in the non-relativistic approximation are given by
\begin{equation}
\dot{x}=\left ( 1+ \frac{\alpha x}{c^2}\right ) \frac{p}{m_0}
~~{\rm{and}}~~ \dot{p}=
-\frac{\alpha}{c^2}\left ( \frac{p^2}{2m_0} + m_0c^2\right )
\end{equation}
On integrating the second one we have
\begin{equation}
p=i2^{1/2} m_0c \cot\left (\frac{2^{1/2}\omega t+\phi}{2}\right )=ip_I
\end{equation}
The particle momentum is therefore purely imaginary in nature with
its real part $p_R=0$. Here $\phi$ is a real constant
phase. Next evaluating the first integral analytically, we have 
\begin{eqnarray}
x&=&\frac{c}{\omega} \left [ -1 +\cos\left \{\ln\left ( \sin^2\left (\frac{2^{1/2}\omega t-\phi}{2}\right )\right ) 
\right \} \right ]
\nonumber \\ &+& i\frac{c}{\omega}\left [ \sin\left \{\ln \left (\sin^2\left (\frac{2^{1/2}\omega t-\phi}{2} \right
) \right ) \right \} \right ]=x_R+ix_I
\end{eqnarray}
The spatial part is therefore complex in nature, where the real part
\begin{equation}
x_R=\frac{c}{\omega} \left [ -1 +\cos\left \{\ln\left ( \sin^2\left (\frac{2^{1/2}\omega t-\phi}{2}\right )\right ) 
\right \} \right ]
\end{equation}
and the corresponding imaginary part is given by
\begin{equation}
x_I=\frac{c}{\omega}\left [ \sin\left \{\ln \left (\sin^2\left (\frac{2^{1/2}\omega t-\phi}{2} \right
) \right ) \right \} \right ]
\end{equation}
Here again eliminating the time part, we have the mathematical form
of phase space trajectories for the imaginary parts only
\begin{equation}
p_I=2^{1/2} m_0c \frac{\left [1-\exp\left \{ \sin^{-1} \left (\frac{\omega}{c}x_I\right ) \right \} \right ]^{1/2}}
{\exp\left\{\frac{1}{2}\sin^{-1}\left ( \frac{\omega}{c}x_I\right )\right \} }
\end{equation}
Which gives the phase space trajectories of the particle in the Rindler space in non-relativistic scenario. It
should be noted here that since the real part of the particle momentum is zero, we have considered the imaginary parts
only. Since $p_I$ is real, therefore $\mid \omega x_I/c\mid \leq 1$, i.e., 
can not have all possible values.

In fig.(2) we have plotted the scaled $x_R$, i.e.  $(\omega x_R/c)$ with
scaled time $(\omega t/2^{1/2})$ for $\phi=0$. Since the constant
phase $\phi$ is completely arbitrary, for the sake of illustration we
have chosen it to be zero. In
this diagram the scaling factors are also taken to be unity. Now if
we consider variation of the scaling factors, the
qualitative nature of the graphs will not change but there will be
quantitative changes.
In fig.(3) we have plotted the scaled $x_I$, i.e.,  $(\omega x_I/c)$ with scaled time 
$(\omega t/2^{1/2})$ for $\phi=0$.  In
this case also same type of changes as has been mentioned for $x_R$ will
 be observed.
In fig.(4) we have plotted the scaled $p_I$, which is actually
$(p_I/2^{1/2}m_0c)$ with scaled time $(\omega t/2^{1/2})$ for $\phi=0$.  In
this case also the scaling factors are exactly equal to one. Further
the same
kind of variation as mentioned above will be observed for $p_I$ with
the change of scaling parameters.
Finally in fig.(5) the phase space trajectory for scaled $x_I$ and scaled $p_I$ is shown
Since the physically accepted domain for scaled $x_I$ is from $-1$ to $0$, we have shown in figs.(6) and (7) the
plot of scaled $x_I$ and scaled $p_I$ with scaled time.

\section{Conclusion}
Finally in conclusion we would like to mention that to the best of our knowledge this is the first time the phase
space trajectories are obtained in Rindler space using non-hermitian $PT$-symmetric Hamiltonian.

In the relativistic case the trajectories can be represented by a set of hyperbolas. Whereas in the non-relativistic
picture, particle momenta are purely imaginary and the space coordinates are complex in nature. The variation of
real and imaginary parts of space coordinates are quite complicated. Further, the phase space is restricted within 
the domain
of negative $x$-values. The imaginary part of  particle momentum has
been observed to change with time in a discrete manner in
this region. 

If we consider the Rindler Hamiltonian in the form
\begin{equation}
H=\left (1+\frac{\alpha x}{c^2}\right )\frac{p^2}{2m}
\end{equation}
then  it is a matter of simple algebra to show that
\begin{equation}
1+\frac{\omega x}{c}= t\exp(2m) ~{\rm{and}}~  p= \frac{2mc}{\omega t}
\end{equation}
Hence redifining $1+\omega x/c$ as new $x$ and $2mc/(\omega p)\exp(2m)$ as new $1/p$, we have
$xp=1$, which gives the phase space trajectories in Rindler space. The trajectories are rectangular hyperbola.

\begin{figure}[ht]
\psfig{figure=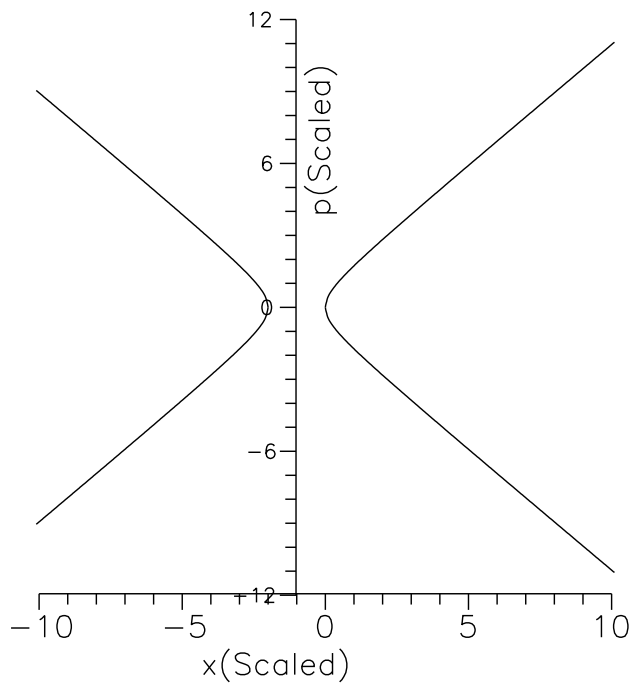,height=0.5\linewidth}
\caption{Phase space trajectories for the relativistic scenario with the scaling parameters equal to unity}
\end{figure}
\begin{figure}[ht]
\psfig{figure=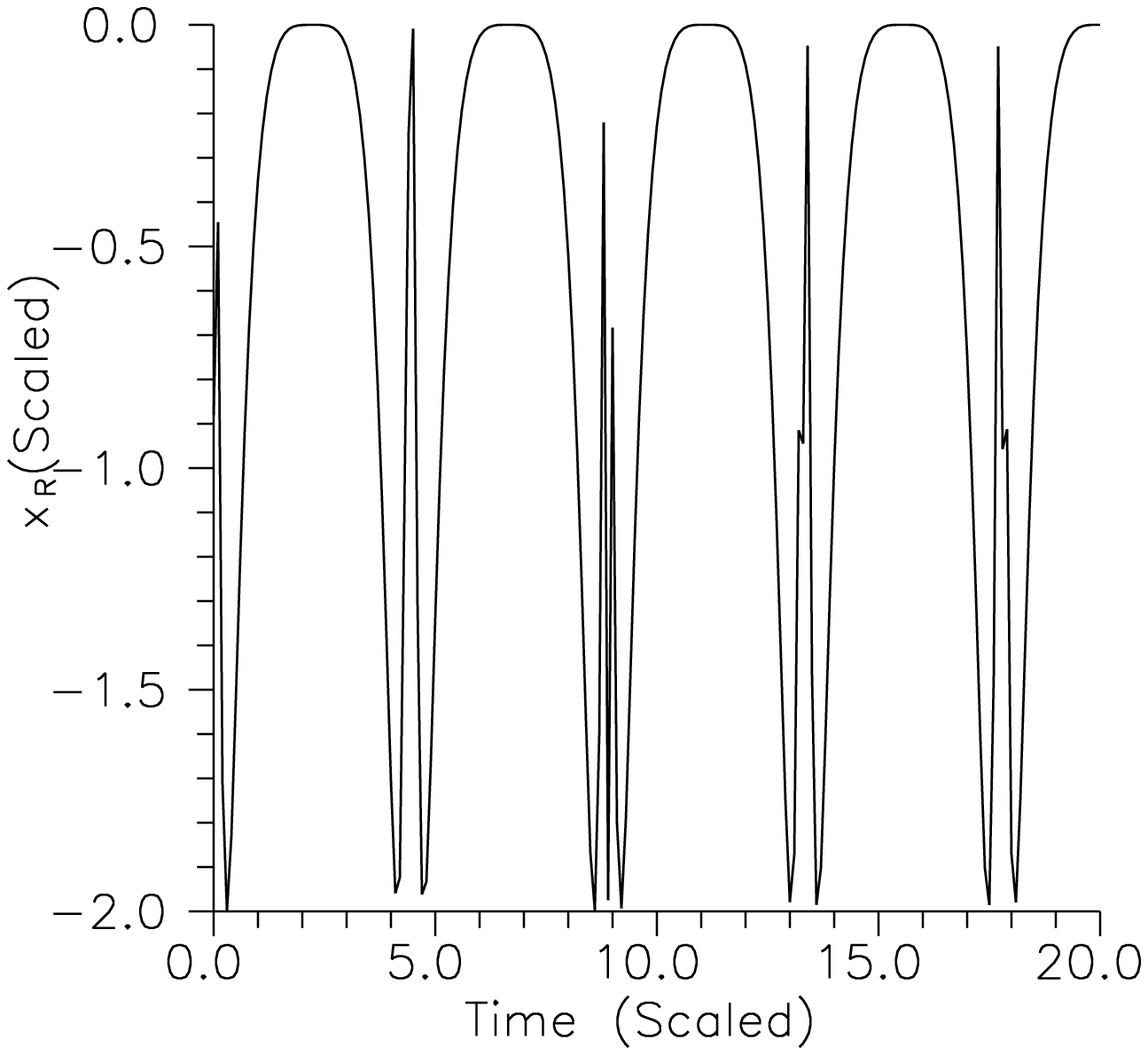,height=0.5\linewidth}
\caption{Variation of scaled $x_R$ with scaled time}
\end{figure}
\begin{figure}[ht]
\psfig{figure=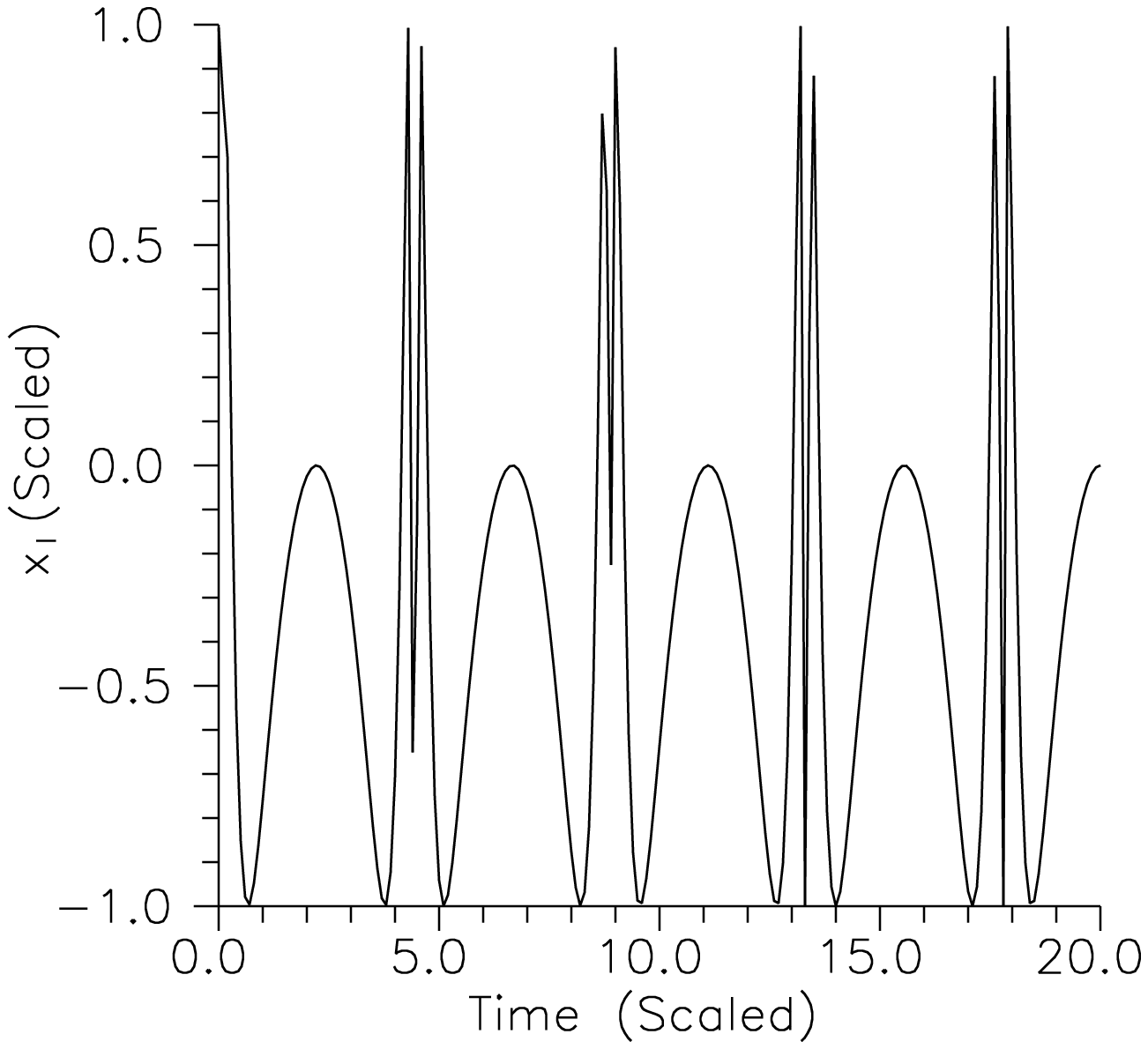,height=0.5\linewidth}
\caption{Variation of scaled $x_I$ with scaled time}
\end{figure}
\begin{figure}[ht]
\psfig{figure=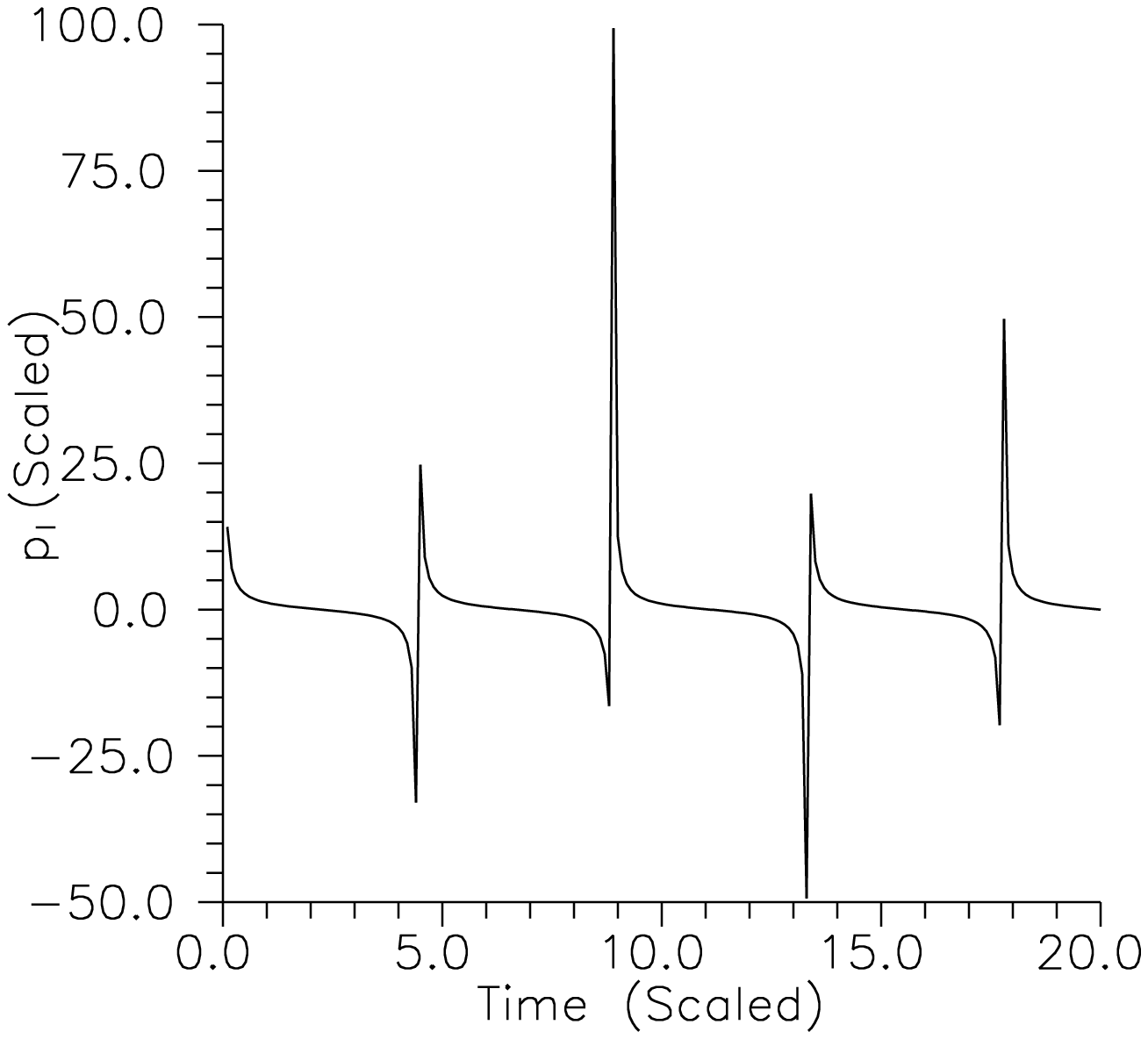,height=0.5\linewidth}
\caption{Variation of scaled $p_I$ with scaled time}
\end{figure}
\begin{figure}[ht]
\psfig{figure=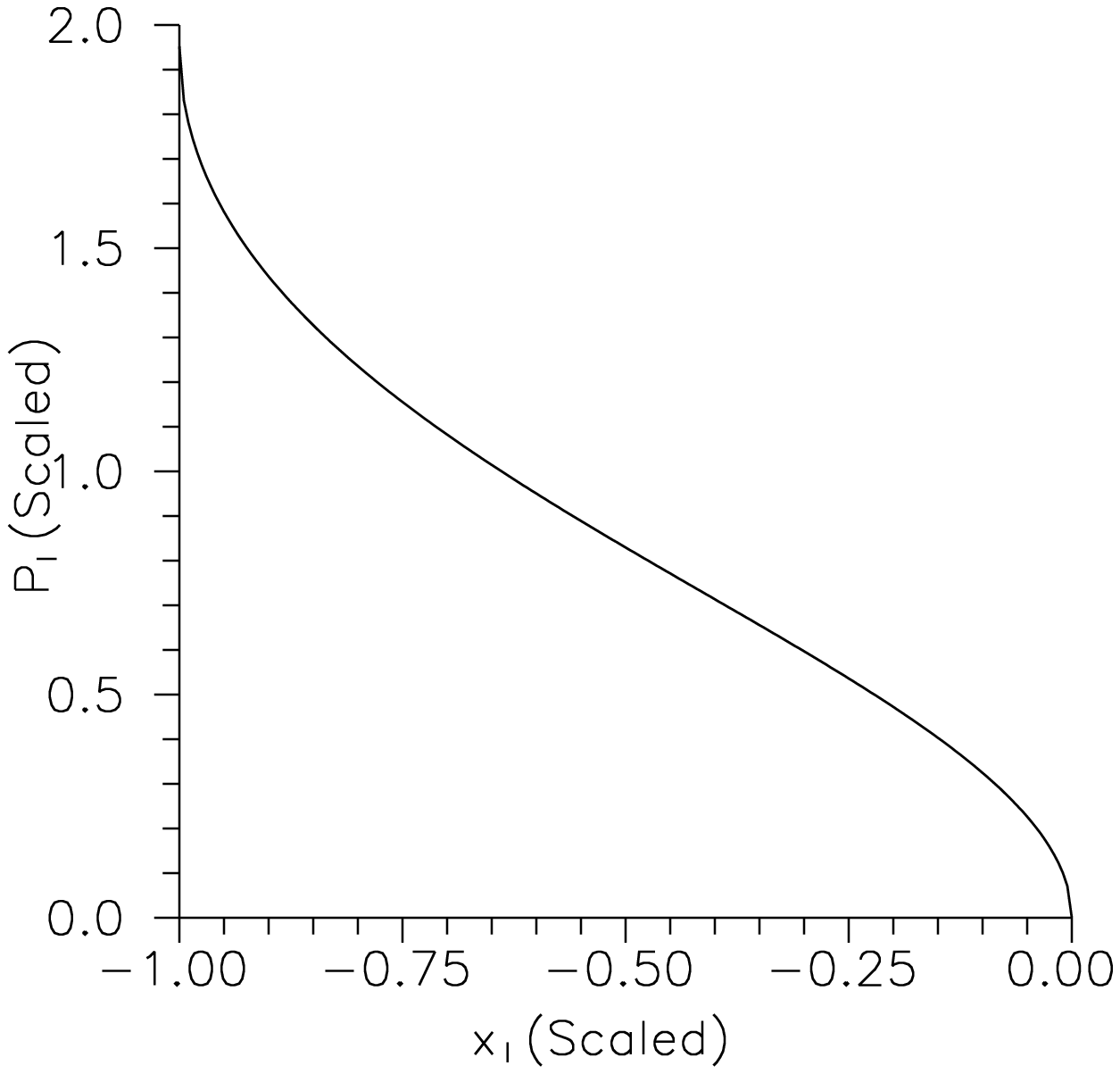,height=0.5\linewidth}
\caption{Phase space trajectories for the non-relativistic scenario
with the scaling parameters equal to unity}
\end{figure}
\begin{figure}[ht]
\psfig{figure=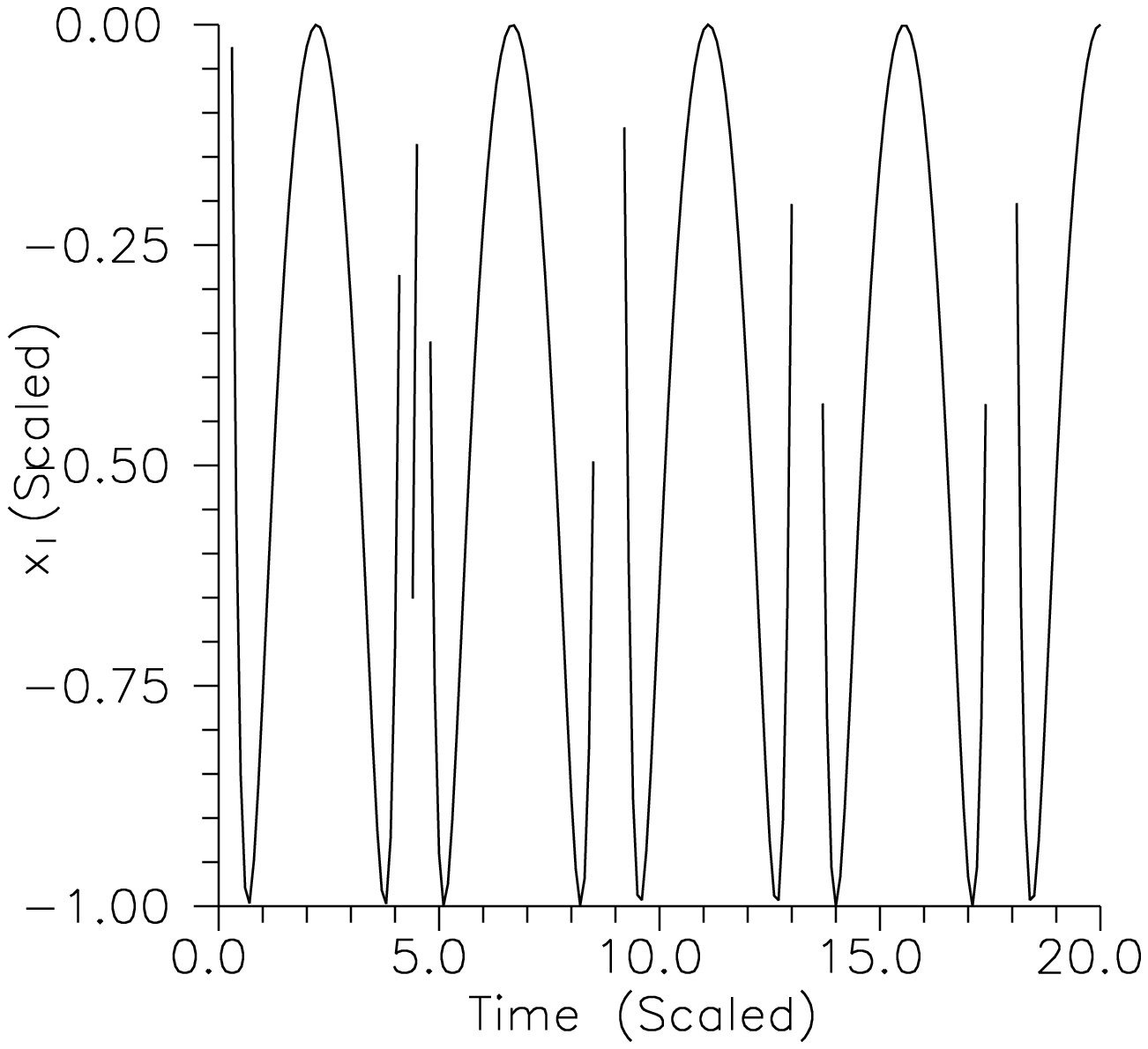,height=0.5\linewidth}
\caption{Temporal variation of $x_I$ in physically acceptable domain}
\end{figure}
\begin{figure}[ht]
\psfig{figure=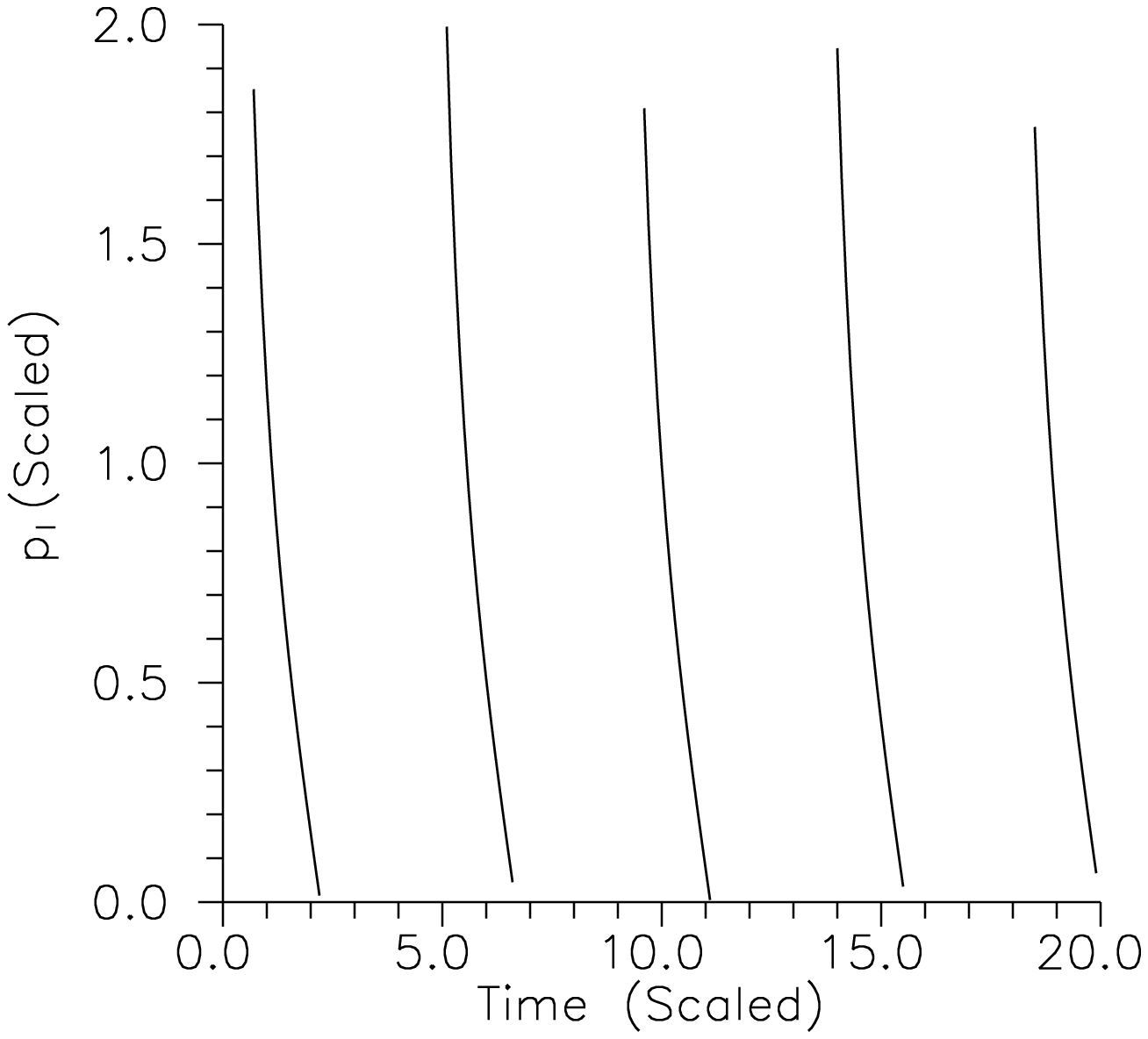,height=0.5\linewidth}
\caption{Temporal variation of $p_I$ in physically acceptable domain}
\end{figure}
\end{document}